\documentclass[aps,prl,reprint,superscriptaddress,amsmath,amssymb,amsfonts,showpacs,floatfix]{revtex4-1}
\usepackage{graphicx}
\usepackage{bm}
\usepackage{calc}
\usepackage{color}
\usepackage{url}
\usepackage{array}
\usepackage{tabularx}

\newif\ifAMStwofonts
\AMStwofontstrue

\def\gsim{~\rlap{$>$}{\lower 1.0ex\hbox{$\sim$}}}

\def\simpropto{\lower.2ex\hbox{$\; \buildrel \propto \over \sim \;$}}
\def\ltsim{\lower.5ex\hbox{$\; \buildrel < \over \sim \;$}}
\def\gtsim{\lower.5ex\hbox{$\; \buildrel > \over \sim \;$}}
\def\ltsim{\lower.5ex\hbox{$\; \buildrel < \over \sim \;$}}
\def\gtsim{\lower.5ex\hbox{$\; \buildrel > \over \sim \;$}}

\def\prd{Physical Review D.}
\def\nat{Nature}

\def\dd{\,{\rm d}}

\def\dd{{\rm d}}

\def\pmb#1{\setbox0=\hbox{#1}%
\kern-.025em\copy0\kern-\wd0
\kern.05em\copy0\kern-\wd0
\kern-.025em\raise.0433em\box0}

\renewcommand{\theenumi}{(\alph{enumi})}

\def\aj{AJ}

\def\simlt{\lower.5ex\hbox{$\; \buildrel < \over \sim \;$}}
\def\simgt{\lower.5ex\hbox{$\; \buildrel > \over \sim \;$}}

\newcommand{\beq}{\begin{equation}}
\newcommand{\eeq}{\end{equation}}
\def\beqa{\begin{eqnarray}}
\def\eeqa{\end{eqnarray}}
\def\fixit#1{}

\def\dd{{\rm d}}
\def\apj{ApJ}
\def\jcap{JCAP}

\def\apjs{ApJ. S}

\def\physrep{Physics Reports}
\def\apjl{ApJL}
\def\mnras{MNRAS}
\def\aap{A\&A}
\def\araa{ARAA}

\begin{document}
\title{Growth rate of cosmological perturbations at \boldmath{$z\sim 0.1$} from a new observational test}

\author{Martin Feix}
\email[Electronic address: ]{mfeix@physics.technion.ac.il}
\affiliation{Department of Physics, Israel Institute of Technology - Technion, Haifa 32000, Israel}
\author{Adi Nusser}
\email[Electronic address: ]{adi@physics.technion.ac.il}
\affiliation{Department of Physics, Israel Institute of Technology - Technion, Haifa 32000, Israel}
\affiliation{Asher Space Science Institute, Israel Institute of Technology - Technion, Haifa 32000, Israel}
\author{Enzo Branchini}
\email[Electronic address: ]{branchin@fis.uniroma3.it}
\affiliation{Department of Physics, Universit\`a Roma Tre, Via della Vasca Navale 84, Rome 00146, Italy}
\affiliation{INFN Sezione di Roma 3, Via della Vasca Navale 84, Rome 00146, Italy}
\affiliation{INAF, Osservatorio Astronomico di Roma, Monte Porzio Catone, Italy}

\begin{abstract}
Spatial variations in the distribution of galaxy luminosities, estimated from redshifts as distance proxies, are
correlated with the peculiar velocity field. Comparing these variations with the peculiar velocities inferred from
galaxy redshift surveys is a powerful test of gravity and dark energy theories on cosmological scales. Using $\sim
2\times 10^{5}$ galaxies from the SDSS Data Release 7, we perform this test in the framework of gravitational instability
to estimate the normalized growth rate of density perturbations $f\sigma_{8} = 0.37\pm 0.13$ at $z\sim 0.1$, which is
in agreement with the $\Lambda$CDM scenario. This unique measurement is complementary to those obtained with more
traditional methods, including clustering analysis. The estimated accuracy at $z\sim 0.1$ is competitive with other
methods when applied to similar datasets.
\end{abstract}

\pacs{98.80.-k, 95.36.+x, 98.62.Py}

\maketitle

\emph{Introduction.}---Unraveling the origin of cosmic acceleration remains one of the biggest challenges in fundamental physics.
Lacking a natural explanation within the standard paradigms of cosmology and particle physics, many theoretical models have been
proposed, ranging from the inclusion of new scalar fields to genuine modifications of general relativity \cite{Frieman2008, Clifton2012}.
Models that provide the same expansion history of the universe can generally lead to a very different evolution of density fluctuations.
In the linear regime, the latter is fully captured by the growth rate,
\begin{equation}
f(\Omega) =  \frac{\dd\log D}{\dd\log a},
\label{eq:1}
\end{equation} 
where $\Omega$ is the matter density parameter, $a$ is the cosmic scale factor, and $D(a)$ is the growing mode of density perturbations.
The standard cold dark matter model with a cosmological constant ($\Lambda$CDM) is well characterized by a growth index
$\gamma=\dd\log f/\dd\log\Omega\approx 6/11$ \cite{Lind05}, but other frameworks typically yield different values and may even exhibit
a scale-dependent behavior \cite[e.g.,][]{Ishak2006, Carroll2006}. Sensitive to both expansion and perturbations, constraining $f$ thus
represents a key observational test to disentangle competing world models, which justifies the increasing number of projects aimed at
its measurement.

Among the most robust probes of $f$ is the analysis of redshift-space distortions (RSDs) \cite{k87, Hamilton1998, Scoccimarro2004}, i.e.
the apparent anisotropic clustering of extragalactic objects such as galaxies \cite{Peacock2001, Guz08, Blake2011, Samushia2014} and quasars
\cite{Angela2008, Ross2009, Mountrichas2009} in redshift space, with an accuracy mainly limited by the uncertainty of modeling nonlinearities
in gravitational dynamics. Methods based on direct measurements of peculiar velocities \cite{DN10, Hudson2012} have also been successful, but
the sparseness and small galaxy numbers in these catalogs restrict them to the very local universe at redshifts $z\sim 0$. Other suggested
approaches include weak lensing, cluster abundances \cite{Weinberg2013}, and Lyman-alpha absorption \cite{Viel2004}.

In this letter, we apply an independent method, based on luminosity modulations, to galaxies from the Sloan Digital Sky Survey Data Release
7 (SDSS DR7) \cite{York2000, abaz}, and obtain a new estimate of the growth rate at $z\sim 0.1$. Altered by the line-of-sight components of
their peculiar motions, $v$, galaxy redshifts generally differ from their actual distances \cite{SW}. Consequently, intrinsic magnitudes $M$
inferred from the observed flux using redshifts appear brighter or dimmer than their true values,
\begin{equation}
M - M^{(t)} = 5\log_{10}\dfrac{D_{L}(z_{c})}{D_{L}(z)},
\label{eq:2}
\end{equation}
where $D_{L}$ is the luminosity distance, $z_{c}$ denotes the cosmological redshift due to the Hubble expansion, and the superscript $(t)$
indicates the true quantities that would be measured if galaxies were correctly placed at $z_{c}$ rather than $z$. For nearby galaxies and
$v/cz\ll 1$ ($c$ is the speed of light), the contribution of velocities to the magnitude shift amounts to $M-M^{(t)}\approx 2.17 v/cz$. On
large scales, the peculiar velocity field is spatially coherent and independently inferable from the observed galaxy distribution in space.
Gravitational instability theory \cite{Peeb80} predicts the velocity field as a function of $\beta=f/b$, where $b$ is the linear bias between
galaxies and the total mass. This prediction allows a correction to $M$, and hence, bounds on $\beta$ can be found by minimizing the scatter
of corrected magnitudes with respect to a reference distribution \cite{Nusser2012}.

\begin{figure*}
\hspace*{\fill}
\includegraphics[width=0.4425\linewidth]{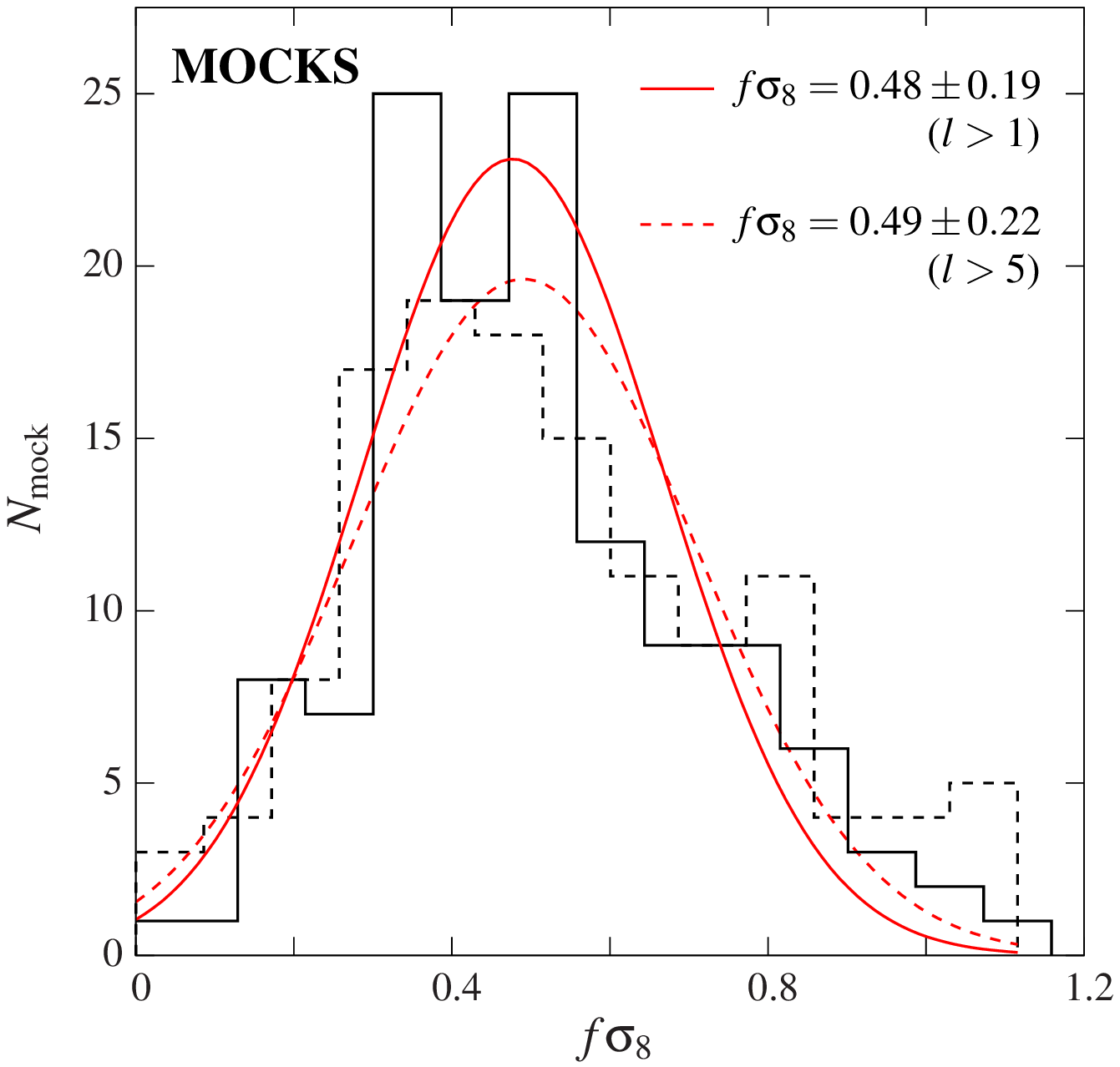}
\hfill
\includegraphics[width=0.4225\linewidth]{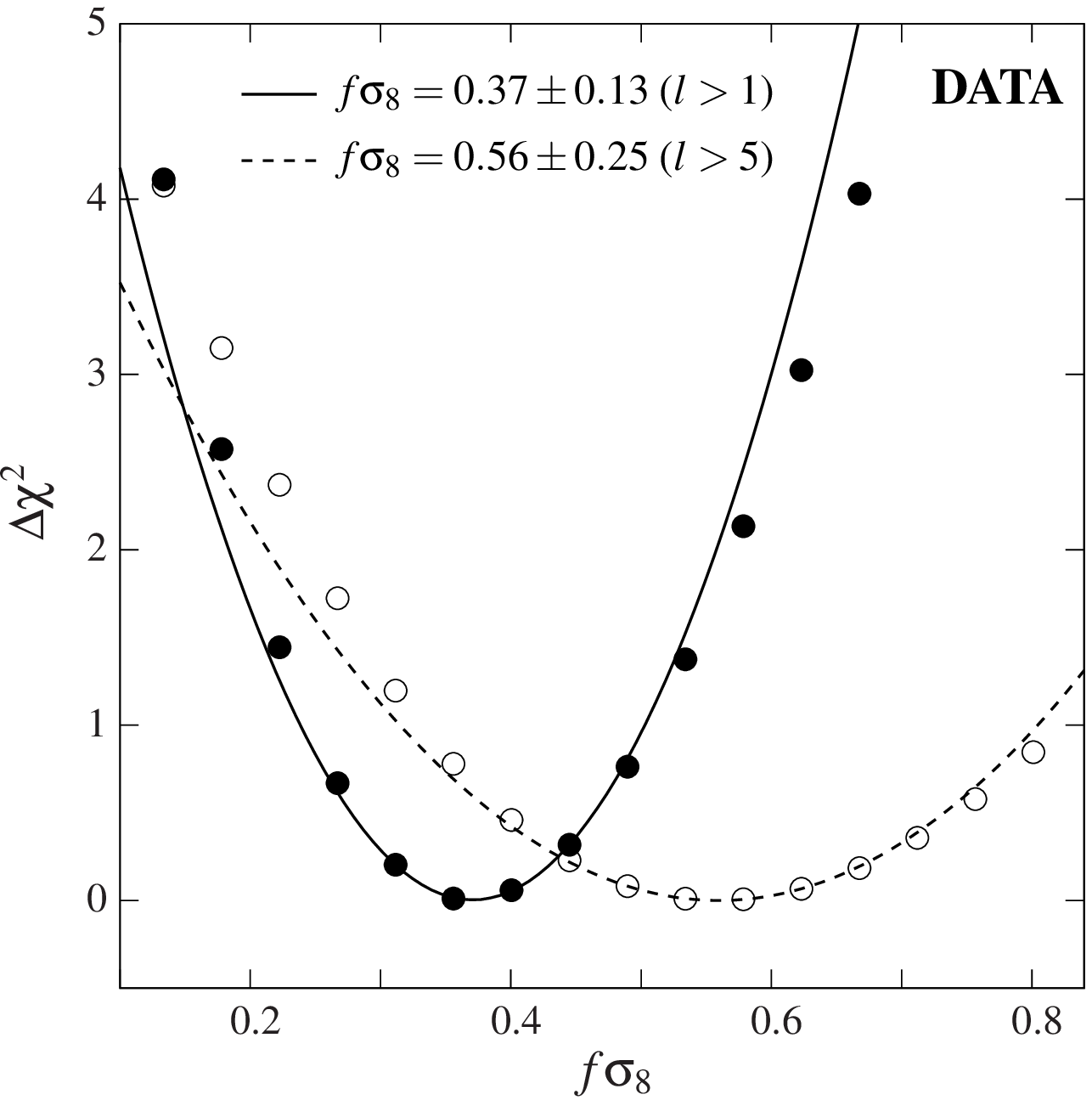}
\hspace*{\fill}
\caption{\emph{Left.}---Histograms of $f\sigma_{8}$-estimates (and Gaussian fits) derived with the luminosity approach from $128$ mock
catalogs, considering only multipoles $l>1$ (solid lines) and $l>5$ (dashed lines) in the velocity reconstructions. The distributions
peak near the true value, $(f\sigma_{8})_{\rm true}\approx 0.44$, and deviate from a symmetric Gaussian mainly because of the velocity
field's nonlinear $\beta$-dependence. \emph{Right.}---Estimated $\Delta\chi^{2}$ (and quadratic approximations) as a function of
$f\sigma_{8}$ for real SDSS galaxies and velocity models with $l>1$ (filled circles) and $l>5$ (open circles). The results assume the
measured power spectrum amplitude of $L^{\star}$-galaxies from Ref. \cite{Tegmark2004}.}
\label{fig1}
\end{figure*}

\emph{Data.}---The analysis is based on the latest version of the NYU Value-Added Galaxy Catalog (NYU-VAGC) derived from the SDSS DR7
\cite{Blanton2005}. To minimize incompleteness and systematics, we focused on the subsample {\tt safe} and considered the same dataset
as in Ref. \cite{Feix2014}. We then selected only galaxies in the range $0.06<z<0.12$ inside a rectangular patch specified in SDSS survey
coordinates, $-33^{\circ}<\eta <36^{\circ}$ and $-48^{\circ}<\lambda <51.5^{\circ}$. This {\tt luminosity} sample, which we used in the
likelihood analysis below, is flux-limited in the $r$-band and contains around $2\times 10^{5}$ galaxies. To build peculiar velocity models,
we also considered a second sample, tagged {\tt velocity}, which corresponds to the maximum volume-limited sample within $0.05<z<0.13$
trimmed to the range $0.06<z<0.12$ and typically includes $\sim 8\times 10^{4}$ galaxies for spatially flat cosmologies with $\Omega\approx
0.3$. To compute absolute magnitudes, we assumed a linear evolution model $Q(z) = 1.6(z-0.1)$ \cite{Feix2014} and $K$-corrections from the
NYU-VAGC \cite{Blanton2007}.

In addition, we used realistic mock SDSS catalogs generated from the Millennium Simulation \cite{Springel2005, Henriques2012} to calibrate
the velocity models and to assess the uncertainty on the $\beta$-estimates. These mocks were customized to match data characteristics such
as number counts, luminosity distribution, and sky coverage.

\emph{Method.}---A full account of the luminosity method can be found in Ref. \cite{Nusser2012}. Here we provide a brief summary of its
key ingredients. Given a galaxy survey with magnitudes, spectroscopic redshifts, and angular positions $\hat{\textbf{\textit{r}}}_{i}$
on the sky, one traces the 3D galaxy distribution and reconstructs the linear peculiar velocity field as a function of $\beta$
\cite{Nusser1994}. The correct value of $\beta$ is then estimated by maximizing the probability of observing the data,
\begin{equation}
\begin{split}
P_{\rm tot} &= \prod\limits_{i}P\left (M_{i}\vert z_{i}, v_{i}(\beta )\right )\\
&= \prod\limits_{i}\left (\phi(M_{i})\middle /\int_{M_{i}^{+}}^{M_{i}^{-}}\phi(M){\rm d}M\right ),
\end{split}
\end{equation}
where $v_{i}(\beta)$ denotes the radial part of the peculiar velocity field evaluated at the position of galaxy $i$, and redshift
errors (typically about $10^{-4}$ for SDSS galaxies, yielding sharp sample cuts in redshift space) are neglected \cite{Nusser2011,
Nusser2012}. Here $\phi(M)$ is the galaxy luminosity function (LF) determined from the full dataset, and the limiting magnitudes
$M^{\pm}$ depend on $v(\beta)$ through the cosmological redshift $z_{c}$. The goal of this approach is to find the $\beta$-value
which minimizes the spread in the observed magnitudes.

Our study does not depend on the Hubble constant and assumes a $\Lambda$CDM cosmology with fixed density parameters taken from Ref.
\cite{Calabrese2013}. Following the procedure of Ref. \cite{Nusser1994}, the velocities were reconstructed in spherical harmonics up
to a multipole $l_{\rm max}=150$ after smoothing the galaxy density field with a Gaussian kernel of $10h^{-1}$ Mpc radius. The monopole
and dipole contributions cannot be reliably modeled for the dataset considered here. Excluding them from the velocity model ($l>1$), we
worked with a full sphere and fixed the boundary conditions by setting the density contrast outside the observed volume to zero. This
particular approach can bias the low-$l$ modes of the velocity model, but the effect is easily estimated and removed with the help of
the mock galaxy catalogs. Since the data covers a narrow $z$-range, we assumed no evolution of $\beta$ and $b$ in the analysis, and computed
a set of model velocity fields varying $\beta$ between 0 and 1 in steps of $\Delta\beta=0.05$. Dominated by large-scale structure, the
cosmic velocity field is insensitive to small-scale features such as those related to galaxy bias. We checked that our models and results
are robust to the precise choice of the smoothing scale.

Assigning $v(\beta)$ to the subsample {\tt luminosity}, we removed galaxies near the sample edges (around 10\%) to avoid artifacts
due to incorrect boundary conditions. To maximize the probability $P_{\rm tot}$, we used a spline-based LF estimator with a separation
$\Delta M=0.5$ \cite{Branchini2012, Feix2014}. The partial sky coverage yields a statistical mixing between the velocity models and
the LF parameters, which may result in biased constraints on $\beta$. To deal with this issue, we set the LF to its estimate for a
vanishing velocity field, evaluated $P_{\rm tot}$ for the different models of $v(\beta)$, and determined the maximum probability through
interpolation. The corresponding bias on $\beta$ was then inferred from the mock catalogs and is at the level of few percent. Although
acceptable in view of the statistical uncertainties, we corrected for this bias in our analysis.

\emph{Results.}---The left panel of Fig. \ref{fig1} shows the distribution of growth rates estimated from 128 mock catalogs for
velocity models with $l>1$. We express our results in terms of $f\sigma_{8}=\beta\sigma_{8}^{\rm gal}$, where $\sigma_{8}^{\rm gal}$
is the measured amplitude of galaxy number counts in spheres of $8h^{-1}$ Mpc radius, related to the amplitude of mass fluctuations
by $\sigma_{8}=\sigma_{8}^{\rm gal}/b$. This combination eliminates $b$ in the comparison to other measurements and cosmological
predictions \cite{Song2009}. In addition to the case $l>1$, we also considered models with $l>5$ to test the robustness to low
multipoles which are most susceptible to the choice of boundary conditions. In both cases, the distributions peak around the true
value, $(f\sigma_{8})_{\rm true}\approx 0.44$, showing an average spread of about 40\%. The results deviate from a symmetric Gaussian
distribution mainly because of the velocity field's nonlinear dependence on $\beta$. Shifting the multipole cutoff from $l>1$ to $l>5$
removes information and slightly increases the observed spread as expected.

The analysis of the real data yielded $\beta=0.42\pm 0.14$ for $l>1$, which, using the measured power spectrum amplitude of $L^{\star}$-galaxies
from Ref. \cite{Tegmark2004}, translates into $f\sigma_{8}=0.37\pm 0.13$ (see right panel of Fig. \ref{fig1}). The quoted errors were derived
from the quadratic approximation of the log-likelihood around its maximum. Similarly, we estimated $\beta=0.63\pm 0.28$ and $f\sigma_{8}=0.56\pm 0.25$
for $l>5$. The corresponding relative errors of 35\% and 45\% are consistent with the mean scatter in Fig. \ref{fig1} and the spread of errors
computed for individual mocks. Both estimates are stable against removing high-$l$ modes ($l>100$) and agree well with the standard $\Lambda$CDM
model which predicts a value of $f\sigma_{8}\approx 0.42$ for our fiducial cosmology. Other choices of cosmological parameters have no significant
impact on these results. Setting $\Omega=1$, for instance, leads to changes of around 13\% for $l>1$ and 4\% for $l>5$.

Performing a suite of basic tests \cite{Feix2014}, we have also verified that our results are insensitive to the adopted $K$-corrections and the
modeling of luminosity evolution. Photometric uncertainties and a possible environmental dependence of the LF play a minor role since the method,
comparing luminosities to peculiar velocities obtained from the very same dataset, is remarkably robust to these effects \cite{Nusser2011, Nusser2012, Feix2014}.

\emph{Conclusions.}---Spatial modulations in the distribution of estimated galaxy luminosities trace the cosmic peculiar velocity field. As
demonstrated by preliminary studies of very local samples \cite{Nusser2012, Branchini2012}, this can be combined with velocity reconstruction
techniques to probe the linear growth factor with galaxy redshift surveys. The modulations in the luminosities of $\sim 2\times 10^{5}$ SDSS
galaxies from the NYU-VAGC yield a value of $f\sigma_{8}$ at $z\sim 0.1$ which is in agreement with the $\Lambda$CDM cosmological model as
dictated by the Planck data \cite{Planck2015}.

\begin{figure} 
\includegraphics[width=0.975\linewidth]{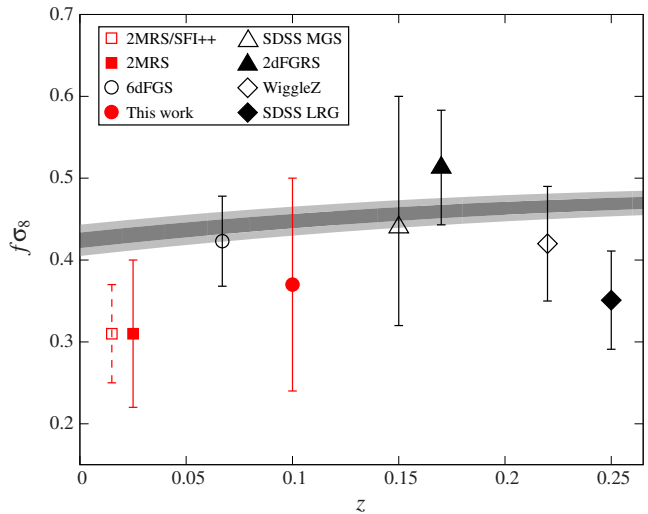}
\caption{Comparison of growth rate measurements at low redshifts. Shown are constraints on $f\sigma_{8}$ from different redshift surveys
based on RSDs \cite{Blake2011, Beutler2012, Samushia2012, Percival2004, Howlett2014}, galaxy luminosities (filled square and circle)
\cite{Nusser2012}, and direct estimates of peculiar velocities (dashed, empty square) \cite{DN10}. All measurements assume a fixed
cosmology and the error bars indicate the respective 68\% confidence intervals. The shaded areas denote the 68\% and 95\% confidence
limits inferred from the Planck data (TT+lowP+lensing) \cite{Planck2015}. The square data points correspond to $z=0.02$ and are offset
for clarity.}
\label{fig2}
\end{figure}

In Fig. \ref{fig2}, we present a compilation of recent measurements of the growth rate from different low-redshift samples ($z<0.3$) which
were obtained from RSDs \cite{Blake2011, Beutler2012, Samushia2012, Percival2004, Howlett2014}, galaxy luminosities (filled square and circle)
\cite{Nusser2012}, and direct estimates of peculiar velocities (dashed, empty square) \cite{DN10}. The consistency between these measurements is
striking in view of the different possible systematic biases associated with the three independent methods. At $z\sim 0.1$, the error in the
estimate of $f\sigma_{8}$ given here is competitive with that of other methods. The accuracy of our estimate (filled circle) matches that from
the RSD analysis of SDSS main galaxies conducted in Ref. \cite{Howlett2014} (empty triangle) which adopts a dataset with similar characteristics.
Methods based on luminosities or distance indicators such as the Tully-Fisher relation \cite{TF77} are less sensitive to nonlinear corrections than
the two-point statistics which enter the analysis of RSDs. Combining these approaches will result in superior control over potential systematics
and allow improved bounds on the growth rate. Further, this will help to resolve the observed trend in RSD measurements of $f\sigma_{8}$ which are
consistently lower than what is expected from Planck \cite{Macaulay2013}. Current and next-generation spectroscopic redshift surveys will feature
larger sky coverage and improved photometric calibration in ground- and space-based experiments \cite[e.g.,][]{Levi2013, euclid2011}. Given these
excellent observational perspectives, we are confident that estimating the growth rate through spatially coherent variations of galaxy luminosities
will be established as a standard method in the context of cosmological data analysis.

\emph{Acknowledgments}---This research was supported by the I-CORE Program of the Planning and Budgeting Committee, THE ISRAEL SCIENCE
FOUNDATION (grants No. 1829/12 and No. 203/09), the Asher Space Research Institute, and the Munich Institute for Astro- and Particle
Physics (MIAPP) of the DFG cluster of excellence ``Origin and Structure of the Universe''. M.F. acknowledges support through a fellowship
from the Minerva Foundation. E.B. is supported by INFN-PD51 INDARK, MIUR PRIN 2011 ``The dark Universe and the cosmic evolution of
baryons: from current surveys to Euclid'', and the Agenzia Spaziale Italiana  from the agreement ASI/INAF/I/023/12/0.

\bibliography{beta_ref.bib}

\begin{thebibliography}{44}%
\makeatletter
\providecommand \@ifxundefined [1]{%
 \@ifx{#1\undefined}
}%
\providecommand \@ifnum [1]{%
 \ifnum #1\expandafter \@firstoftwo
 \else \expandafter \@secondoftwo
 \fi
}%
\providecommand \@ifx [1]{%
 \ifx #1\expandafter \@firstoftwo
 \else \expandafter \@secondoftwo
 \fi
}%
\providecommand \natexlab [1]{#1}%
\providecommand \enquote  [1]{``#1''}%
\providecommand \bibnamefont  [1]{#1}%
\providecommand \bibfnamefont [1]{#1}%
\providecommand \citenamefont [1]{#1}%
\providecommand \href@noop [0]{\@secondoftwo}%
\providecommand \href [0]{\begingroup \@sanitize@url \@href}%
\providecommand \@href[1]{\@@startlink{#1}\@@href}%
\providecommand \@@href[1]{\endgroup#1\@@endlink}%
\providecommand \@sanitize@url [0]{\catcode `\\12\catcode `\$12\catcode
  `\&12\catcode `\#12\catcode `\^12\catcode `\_12\catcode `\%12\relax}%
\providecommand \@@startlink[1]{}%
\providecommand \@@endlink[0]{}%
\providecommand \url  [0]{\begingroup\@sanitize@url \@url }%
\providecommand \@url [1]{\endgroup\@href {#1}{\urlprefix }}%
\providecommand \urlprefix  [0]{URL }%
\providecommand \Eprint [0]{\href }%
\@ifxundefined \urlstyle {%
  \providecommand \doi  [0]{\begingroup \@sanitize@url \@doi}%
  \providecommand \@doi [1]{\endgroup \@@startlink {\doibase
  #1}doi:\discretionary {}{}{}#1\@@endlink }%
}{%
  \providecommand \doi  [0]{doi:\discretionary{}{}{}\begingroup
  \urlstyle{rm}\Url }%
}%
\providecommand \doibase [0]{http://dx.doi.org/}%
\providecommand \Doi [0]{\begingroup \@sanitize@url \@Doi }%
\providecommand \@Doi  [1]{\endgroup\@@startlink{\doibase#1}\@@Doi}%
\providecommand \@@Doi [1]{#1\@@endlink}%
\providecommand \selectlanguage [0]{\@gobble}%
\providecommand \bibinfo  [0]{\@secondoftwo}%
\providecommand \bibfield  [0]{\@secondoftwo}%
\providecommand \translation [1]{[#1]}%
\providecommand \BibitemOpen [0]{}%
\providecommand \bibitemStop [0]{}%
\providecommand \bibitemNoStop [0]{.\EOS\space}%
\providecommand \EOS [0]{\spacefactor3000\relax}%
\providecommand \BibitemShut  [1]{\csname bibitem#1\endcsname}%
\bibitem [{\citenamefont {{Frieman}}\ \emph {et~al.}(2008)\citenamefont
  {{Frieman}}, \citenamefont {{Turner}},\ and\ \citenamefont
  {{Huterer}}}]{Frieman2008}%
  \BibitemOpen
  \bibfield  {author} {\bibinfo {author} {\bibfnamefont {J.~A.}\ \bibnamefont
  {{Frieman}}}, \bibinfo {author} {\bibfnamefont {M.~S.}\ \bibnamefont
  {{Turner}}}, \ and\ \bibinfo {author} {\bibfnamefont {D.}~\bibnamefont
  {{Huterer}}},\ }\Doi {10.1146/annurev.astro.46.060407.145243} {\bibfield
  {journal} {\bibinfo  {journal} {\araa},\ }\textbf {\bibinfo {volume} {46}},\
  \bibinfo {pages} {385} (\bibinfo {year} {2008})},\ \Eprint
  {http://arxiv.org/abs/0803.0982} {arXiv:0803.0982} \BibitemShut {NoStop}%
\bibitem [{\citenamefont {{Clifton}}\ \emph {et~al.}(2012)\citenamefont
  {{Clifton}}, \citenamefont {{Ferreira}}, \citenamefont {{Padilla}},\ and\
  \citenamefont {{Skordis}}}]{Clifton2012}%
  \BibitemOpen
  \bibfield  {author} {\bibinfo {author} {\bibfnamefont {T.}~\bibnamefont
  {{Clifton}}}, \bibinfo {author} {\bibfnamefont {P.~G.}\ \bibnamefont
  {{Ferreira}}}, \bibinfo {author} {\bibfnamefont {A.}~\bibnamefont
  {{Padilla}}}, \ and\ \bibinfo {author} {\bibfnamefont {C.}~\bibnamefont
  {{Skordis}}},\ }\Doi {10.1016/j.physrep.2012.01.001} {\bibfield  {journal}
  {\bibinfo  {journal} {\physrep},\ }\textbf {\bibinfo {volume} {513}},\
  \bibinfo {pages} {1} (\bibinfo {year} {2012})},\ \Eprint
  {http://arxiv.org/abs/1106.2476} {arXiv:1106.2476 [astro-ph.CO]} \BibitemShut
  {NoStop}%
\bibitem [{\citenamefont {{Linder}}(2005)}]{Lind05}%
  \BibitemOpen
  \bibfield  {author} {\bibinfo {author} {\bibfnamefont {E.~V.}\ \bibnamefont
  {{Linder}}},\ }\Doi {10.1103/PhysRevD.72.043529} {\bibfield  {journal}
  {\bibinfo  {journal} {\prd},\ }\textbf {\bibinfo {volume} {72}},\ \bibinfo
  {pages} {043529} (\bibinfo {year} {2005})},\ \Eprint
  {http://arxiv.org/abs/arXiv:astro-ph/0507263} {arXiv:astro-ph/0507263}
  \BibitemShut {NoStop}%
\bibitem [{\citenamefont {{Ishak}}\ \emph {et~al.}(2006)\citenamefont
  {{Ishak}}, \citenamefont {{Upadhye}},\ and\ \citenamefont
  {{Spergel}}}]{Ishak2006}%
  \BibitemOpen
  \bibfield  {author} {\bibinfo {author} {\bibfnamefont {M.}~\bibnamefont
  {{Ishak}}}, \bibinfo {author} {\bibfnamefont {A.}~\bibnamefont {{Upadhye}}},
  \ and\ \bibinfo {author} {\bibfnamefont {D.~N.}\ \bibnamefont {{Spergel}}},\
  }\Doi {10.1103/PhysRevD.74.043513} {\bibfield  {journal} {\bibinfo  {journal}
  {\prd},\ }\textbf {\bibinfo {volume} {74}},\ \bibinfo {eid} {043513}
  (\bibinfo {year} {2006})},\ \Eprint {http://arxiv.org/abs/astro-ph/0507184}
  {astro-ph/0507184} \BibitemShut {NoStop}%
\bibitem [{\citenamefont {{Carroll}}\ \emph {et~al.}(2006)\citenamefont
  {{Carroll}}, \citenamefont {{Sawicki}}, \citenamefont {{Silvestri}},\ and\
  \citenamefont {{Trodden}}}]{Carroll2006}%
  \BibitemOpen
  \bibfield  {author} {\bibinfo {author} {\bibfnamefont {S.~M.}\ \bibnamefont
  {{Carroll}}}, \bibinfo {author} {\bibfnamefont {I.}~\bibnamefont
  {{Sawicki}}}, \bibinfo {author} {\bibfnamefont {A.}~\bibnamefont
  {{Silvestri}}}, \ and\ \bibinfo {author} {\bibfnamefont {M.}~\bibnamefont
  {{Trodden}}},\ }\Doi {10.1088/1367-2630/8/12/323} {\bibfield  {journal}
  {\bibinfo  {journal} {New Journal of Physics},\ }\textbf {\bibinfo {volume}
  {8}},\ \bibinfo {pages} {323} (\bibinfo {year} {2006})},\ \Eprint
  {http://arxiv.org/abs/astro-ph/0607458} {astro-ph/0607458} \BibitemShut
  {NoStop}%
\bibitem [{\citenamefont {{Kaiser}}(1987)}]{k87}%
  \BibitemOpen
  \bibfield  {author} {\bibinfo {author} {\bibfnamefont {N.}~\bibnamefont
  {{Kaiser}}},\ }\href@noop {} {\bibfield  {journal} {\bibinfo  {journal}
  {\mnras},\ }\textbf {\bibinfo {volume} {227}},\ \bibinfo {pages} {1}
  (\bibinfo {year} {1987})}\BibitemShut {NoStop}%
\bibitem [{\citenamefont {{Hamilton}}(1998)}]{Hamilton1998}%
  \BibitemOpen
  \bibfield  {author} {\bibinfo {author} {\bibfnamefont {A.~J.~S.}\
  \bibnamefont {{Hamilton}}},\ }in\ \href@noop {} {\emph {\bibinfo {booktitle}
  {The Evolving Universe}}},\ \bibinfo {series} {Astrophysics and Space Science
  Library}, Vol.\ \bibinfo {volume} {231},\ \bibinfo {editor} {edited by\
  \bibinfo {editor} {\bibfnamefont {D.}~\bibnamefont {{Hamilton}}}}\ (\bibinfo
  {year} {1998})\ p.\ \bibinfo {pages} {185}\BibitemShut {NoStop}%
\bibitem [{\citenamefont {{Scoccimarro}}(2004)}]{Scoccimarro2004}%
  \BibitemOpen
  \bibfield  {author} {\bibinfo {author} {\bibfnamefont {R.}~\bibnamefont
  {{Scoccimarro}}},\ }\Doi {10.1103/PhysRevD.70.083007} {\bibfield  {journal}
  {\bibinfo  {journal} {\prd},\ }\textbf {\bibinfo {volume} {70}},\ \bibinfo
  {eid} {083007} (\bibinfo {year} {2004})},\ \Eprint
  {http://arxiv.org/abs/astro-ph/0407214} {astro-ph/0407214} \BibitemShut
  {NoStop}%
\bibitem [{\citenamefont {{Peacock}}\ \emph {et~al.}(2001)\citenamefont
  {{Peacock}}, \citenamefont {{Cole}}, \citenamefont {{Norberg}}, \citenamefont
  {{Baugh}}, \citenamefont {{Bland-Hawthorn}}, \citenamefont {{Bridges}},\ and\
  \citenamefont {et~al.}}]{Peacock2001}%
  \BibitemOpen
  \bibfield  {author} {\bibinfo {author} {\bibfnamefont {J.~A.}\ \bibnamefont
  {{Peacock}}}, \bibinfo {author} {\bibfnamefont {S.}~\bibnamefont {{Cole}}},
  \bibinfo {author} {\bibfnamefont {P.}~\bibnamefont {{Norberg}}}, \bibinfo
  {author} {\bibfnamefont {C.~M.}\ \bibnamefont {{Baugh}}}, \bibinfo {author}
  {\bibfnamefont {J.}~\bibnamefont {{Bland-Hawthorn}}}, \bibinfo {author}
  {\bibfnamefont {T.}~\bibnamefont {{Bridges}}}, \ and\ \bibinfo {author}
  {\bibnamefont {et~al.}},\ }\href@noop {} {\bibfield  {journal} {\bibinfo
  {journal} {\nat},\ }\textbf {\bibinfo {volume} {410}},\ \bibinfo {pages}
  {169} (\bibinfo {year} {2001})},\ \Eprint
  {http://arxiv.org/abs/astro-ph/0103143} {astro-ph/0103143} \BibitemShut
  {NoStop}%
\bibitem [{\citenamefont {{Guzzo}}\ \emph {et~al.}(2008)\citenamefont
  {{Guzzo}}, \citenamefont {{Pierleoni}}, \citenamefont {{Meneux}},
  \citenamefont {{Branchini}}, \citenamefont {{Le F{\`e}vre}}, \citenamefont
  {{Marinoni}},\ and\ \citenamefont {et~al.}}]{Guz08}%
  \BibitemOpen
  \bibfield  {author} {\bibinfo {author} {\bibfnamefont {L.}~\bibnamefont
  {{Guzzo}}}, \bibinfo {author} {\bibfnamefont {M.}~\bibnamefont
  {{Pierleoni}}}, \bibinfo {author} {\bibfnamefont {B.}~\bibnamefont
  {{Meneux}}}, \bibinfo {author} {\bibfnamefont {E.}~\bibnamefont
  {{Branchini}}}, \bibinfo {author} {\bibfnamefont {O.}~\bibnamefont {{Le
  F{\`e}vre}}}, \bibinfo {author} {\bibfnamefont {C.}~\bibnamefont
  {{Marinoni}}}, \ and\ \bibinfo {author} {\bibnamefont {et~al.}},\ }\Doi
  {10.1038/nature06555} {\bibfield  {journal} {\bibinfo  {journal} {\nat},\
  }\textbf {\bibinfo {volume} {451}},\ \bibinfo {pages} {541} (\bibinfo {year}
  {2008})},\ \Eprint {http://arxiv.org/abs/0802.1944} {arXiv:0802.1944}
  \BibitemShut {NoStop}%
\bibitem [{\citenamefont {{Blake}}\ \emph {et~al.}(2011)\citenamefont
  {{Blake}}, \citenamefont {{Brough}}, \citenamefont {{Colless}}, \citenamefont
  {{Contreras}}, \citenamefont {{Couch}}, \citenamefont {{Croom}},\ and\
  \citenamefont {et~al.}}]{Blake2011}%
  \BibitemOpen
  \bibfield  {author} {\bibinfo {author} {\bibfnamefont {C.}~\bibnamefont
  {{Blake}}}, \bibinfo {author} {\bibfnamefont {S.}~\bibnamefont {{Brough}}},
  \bibinfo {author} {\bibfnamefont {M.}~\bibnamefont {{Colless}}}, \bibinfo
  {author} {\bibfnamefont {C.}~\bibnamefont {{Contreras}}}, \bibinfo {author}
  {\bibfnamefont {W.}~\bibnamefont {{Couch}}}, \bibinfo {author} {\bibfnamefont
  {S.}~\bibnamefont {{Croom}}}, \ and\ \bibinfo {author} {\bibnamefont
  {et~al.}},\ }\Doi {10.1111/j.1365-2966.2011.18903.x} {\bibfield  {journal}
  {\bibinfo  {journal} {\mnras},\ \bibinfo {pages} {834}} (\bibinfo {year}
  {2011})},\ \Eprint {http://arxiv.org/abs/1104.2948} {arXiv:1104.2948
  [astro-ph.CO]} \BibitemShut {NoStop}%
\bibitem [{\citenamefont {{Samushia}}\ \emph {et~al.}(2014)\citenamefont
  {{Samushia}}, \citenamefont {{Reid}}, \citenamefont {{White}}, \citenamefont
  {{Percival}}, \citenamefont {{Cuesta}}, \citenamefont {{Zhao}},\ and\
  \citenamefont {et~al.}}]{Samushia2014}%
  \BibitemOpen
  \bibfield  {author} {\bibinfo {author} {\bibfnamefont {L.}~\bibnamefont
  {{Samushia}}}, \bibinfo {author} {\bibfnamefont {B.~A.}\ \bibnamefont
  {{Reid}}}, \bibinfo {author} {\bibfnamefont {M.}~\bibnamefont {{White}}},
  \bibinfo {author} {\bibfnamefont {W.~J.}\ \bibnamefont {{Percival}}},
  \bibinfo {author} {\bibfnamefont {A.~J.}\ \bibnamefont {{Cuesta}}}, \bibinfo
  {author} {\bibfnamefont {G.-B.}\ \bibnamefont {{Zhao}}}, \ and\ \bibinfo
  {author} {\bibnamefont {et~al.}},\ }\Doi {10.1093/mnras/stu197} {\bibfield
  {journal} {\bibinfo  {journal} {\mnras},\ }\textbf {\bibinfo {volume}
  {439}},\ \bibinfo {pages} {3504} (\bibinfo {year} {2014})},\ \Eprint
  {http://arxiv.org/abs/1312.4899} {arXiv:1312.4899} \BibitemShut {NoStop}%
\bibitem [{\citenamefont {{da {\^A}ngela}}\ \emph {et~al.}(2008)\citenamefont
  {{da {\^A}ngela}}, \citenamefont {{Shanks}}, \citenamefont {{Croom}},
  \citenamefont {{Weilbacher}}, \citenamefont {{Brunner}}, \citenamefont
  {{Couch}},\ and\ \citenamefont {et~al.}}]{Angela2008}%
  \BibitemOpen
  \bibfield  {author} {\bibinfo {author} {\bibfnamefont {J.}~\bibnamefont {{da
  {\^A}ngela}}}, \bibinfo {author} {\bibfnamefont {T.}~\bibnamefont
  {{Shanks}}}, \bibinfo {author} {\bibfnamefont {S.~M.}\ \bibnamefont
  {{Croom}}}, \bibinfo {author} {\bibfnamefont {P.}~\bibnamefont
  {{Weilbacher}}}, \bibinfo {author} {\bibfnamefont {R.~J.}\ \bibnamefont
  {{Brunner}}}, \bibinfo {author} {\bibfnamefont {W.~J.}\ \bibnamefont
  {{Couch}}}, \ and\ \bibinfo {author} {\bibnamefont {et~al.}},\ }\Doi
  {10.1111/j.1365-2966.2007.12552.x} {\bibfield  {journal} {\bibinfo  {journal}
  {\mnras},\ }\textbf {\bibinfo {volume} {383}},\ \bibinfo {pages} {565}
  (\bibinfo {year} {2008})},\ \Eprint {http://arxiv.org/abs/astro-ph/0612401}
  {astro-ph/0612401} \BibitemShut {NoStop}%
\bibitem [{\citenamefont {{Ross}}\ \emph {et~al.}(2009)\citenamefont {{Ross}},
  \citenamefont {{Shen}}, \citenamefont {{Strauss}}, \citenamefont {{Vanden
  Berk}}, \citenamefont {{Connolly}}, \citenamefont {{Richards}},\ and\
  \citenamefont {et~al.}}]{Ross2009}%
  \BibitemOpen
  \bibfield  {author} {\bibinfo {author} {\bibfnamefont {N.~P.}\ \bibnamefont
  {{Ross}}}, \bibinfo {author} {\bibfnamefont {Y.}~\bibnamefont {{Shen}}},
  \bibinfo {author} {\bibfnamefont {M.~A.}\ \bibnamefont {{Strauss}}}, \bibinfo
  {author} {\bibfnamefont {D.~E.}\ \bibnamefont {{Vanden Berk}}}, \bibinfo
  {author} {\bibfnamefont {A.~J.}\ \bibnamefont {{Connolly}}}, \bibinfo
  {author} {\bibfnamefont {G.~T.}\ \bibnamefont {{Richards}}}, \ and\ \bibinfo
  {author} {\bibnamefont {et~al.}},\ }\Doi {10.1088/0004-637X/697/2/1634}
  {\bibfield  {journal} {\bibinfo  {journal} {\apj},\ }\textbf {\bibinfo
  {volume} {697}},\ \bibinfo {pages} {1634} (\bibinfo {year} {2009})},\ \Eprint
  {http://arxiv.org/abs/0903.3230} {arXiv:0903.3230 [astro-ph.CO]} \BibitemShut
  {NoStop}%
\bibitem [{\citenamefont {{Mountrichas}}\ \emph {et~al.}(2009)\citenamefont
  {{Mountrichas}}, \citenamefont {{Sawangwit}}, \citenamefont {{Shanks}},
  \citenamefont {{Croom}}, \citenamefont {{Schneider}}, \citenamefont
  {{Myers}},\ and\ \citenamefont {{Pimbblet}}}]{Mountrichas2009}%
  \BibitemOpen
  \bibfield  {author} {\bibinfo {author} {\bibfnamefont {G.}~\bibnamefont
  {{Mountrichas}}}, \bibinfo {author} {\bibfnamefont {U.}~\bibnamefont
  {{Sawangwit}}}, \bibinfo {author} {\bibfnamefont {T.}~\bibnamefont
  {{Shanks}}}, \bibinfo {author} {\bibfnamefont {S.~M.}\ \bibnamefont
  {{Croom}}}, \bibinfo {author} {\bibfnamefont {D.~P.}\ \bibnamefont
  {{Schneider}}}, \bibinfo {author} {\bibfnamefont {A.~D.}\ \bibnamefont
  {{Myers}}}, \ and\ \bibinfo {author} {\bibfnamefont {K.}~\bibnamefont
  {{Pimbblet}}},\ }\Doi {10.1111/j.1365-2966.2009.14456.x} {\bibfield
  {journal} {\bibinfo  {journal} {\mnras},\ }\textbf {\bibinfo {volume}
  {394}},\ \bibinfo {pages} {2050} (\bibinfo {year} {2009})},\ \Eprint
  {http://arxiv.org/abs/0801.1816} {arXiv:0801.1816} \BibitemShut {NoStop}%
\bibitem [{\citenamefont {{Davis}}\ \emph {et~al.}(2011)\citenamefont
  {{Davis}}, \citenamefont {{Nusser}}, \citenamefont {{Masters}}, \citenamefont
  {{Springob}}, \citenamefont {{Huchra}},\ and\ \citenamefont
  {{Lemson}}}]{DN10}%
  \BibitemOpen
  \bibfield  {author} {\bibinfo {author} {\bibfnamefont {M.}~\bibnamefont
  {{Davis}}}, \bibinfo {author} {\bibfnamefont {A.}~\bibnamefont {{Nusser}}},
  \bibinfo {author} {\bibfnamefont {K.~L.}\ \bibnamefont {{Masters}}}, \bibinfo
  {author} {\bibfnamefont {C.}~\bibnamefont {{Springob}}}, \bibinfo {author}
  {\bibfnamefont {J.~P.}\ \bibnamefont {{Huchra}}}, \ and\ \bibinfo {author}
  {\bibfnamefont {G.}~\bibnamefont {{Lemson}}},\ }\Doi
  {10.1111/j.1365-2966.2011.18362.x} {\bibfield  {journal} {\bibinfo  {journal}
  {\mnras},\ }\textbf {\bibinfo {volume} {413}},\ \bibinfo {pages} {2906}
  (\bibinfo {year} {2011})},\ \Eprint {http://arxiv.org/abs/1011.3114}
  {arXiv:1011.3114 [astro-ph.CO]} \BibitemShut {NoStop}%
\bibitem [{\citenamefont {{Hudson}}\ and\ \citenamefont
  {{Turnbull}}(2012)}]{Hudson2012}%
  \BibitemOpen
  \bibfield  {author} {\bibinfo {author} {\bibfnamefont {M.~J.}\ \bibnamefont
  {{Hudson}}}\ and\ \bibinfo {author} {\bibfnamefont {S.~J.}\ \bibnamefont
  {{Turnbull}}},\ }\Doi {10.1088/2041-8205/751/2/L30} {\bibfield  {journal}
  {\bibinfo  {journal} {\apjl},\ }\textbf {\bibinfo {volume} {751}},\ \bibinfo
  {eid} {L30} (\bibinfo {year} {2012})},\ \Eprint
  {http://arxiv.org/abs/1203.4814} {arXiv:1203.4814 [astro-ph.CO]} \BibitemShut
  {NoStop}%
\bibitem [{\citenamefont {{Weinberg}}\ \emph {et~al.}(2013)\citenamefont
  {{Weinberg}}, \citenamefont {{Mortonson}}, \citenamefont {{Eisenstein}},
  \citenamefont {{Hirata}}, \citenamefont {{Riess}},\ and\ \citenamefont
  {{Rozo}}}]{Weinberg2013}%
  \BibitemOpen
  \bibfield  {author} {\bibinfo {author} {\bibfnamefont {D.~H.}\ \bibnamefont
  {{Weinberg}}}, \bibinfo {author} {\bibfnamefont {M.~J.}\ \bibnamefont
  {{Mortonson}}}, \bibinfo {author} {\bibfnamefont {D.~J.}\ \bibnamefont
  {{Eisenstein}}}, \bibinfo {author} {\bibfnamefont {C.}~\bibnamefont
  {{Hirata}}}, \bibinfo {author} {\bibfnamefont {A.~G.}\ \bibnamefont
  {{Riess}}}, \ and\ \bibinfo {author} {\bibfnamefont {E.}~\bibnamefont
  {{Rozo}}},\ }\Doi {10.1016/j.physrep.2013.05.001} {\bibfield  {journal}
  {\bibinfo  {journal} {\physrep},\ }\textbf {\bibinfo {volume} {530}},\
  \bibinfo {pages} {87} (\bibinfo {year} {2013})},\ \Eprint
  {http://arxiv.org/abs/1201.2434} {arXiv:1201.2434 [astro-ph.CO]} \BibitemShut
  {NoStop}%
\bibitem [{\citenamefont {{Viel}}\ \emph {et~al.}(2004)\citenamefont {{Viel}},
  \citenamefont {{Haehnelt}},\ and\ \citenamefont {{Springel}}}]{Viel2004}%
  \BibitemOpen
  \bibfield  {author} {\bibinfo {author} {\bibfnamefont {M.}~\bibnamefont
  {{Viel}}}, \bibinfo {author} {\bibfnamefont {M.~G.}\ \bibnamefont
  {{Haehnelt}}}, \ and\ \bibinfo {author} {\bibfnamefont {V.}~\bibnamefont
  {{Springel}}},\ }\Doi {10.1111/j.1365-2966.2004.08224.x} {\bibfield
  {journal} {\bibinfo  {journal} {\mnras},\ }\textbf {\bibinfo {volume}
  {354}},\ \bibinfo {pages} {684} (\bibinfo {year} {2004})},\ \Eprint
  {http://arxiv.org/abs/astro-ph/0404600} {astro-ph/0404600} \BibitemShut
  {NoStop}%
\bibitem [{\citenamefont {{York}}\ \emph {et~al.}(2000)\citenamefont {{York}},
  \citenamefont {{Adelman}}, \citenamefont {{Anderson}}, \citenamefont
  {{Anderson}}, \citenamefont {{Annis}}, \citenamefont {{Bahcall}},\ and\
  \citenamefont {et~al.}}]{York2000}%
  \BibitemOpen
  \bibfield  {author} {\bibinfo {author} {\bibfnamefont {D.~G.}\ \bibnamefont
  {{York}}}, \bibinfo {author} {\bibfnamefont {J.}~\bibnamefont {{Adelman}}},
  \bibinfo {author} {\bibfnamefont {J.~E.}\ \bibnamefont {{Anderson}},
  \bibfnamefont {Jr.}}, \bibinfo {author} {\bibfnamefont {S.~F.}\ \bibnamefont
  {{Anderson}}}, \bibinfo {author} {\bibfnamefont {J.}~\bibnamefont {{Annis}}},
  \bibinfo {author} {\bibfnamefont {N.~A.}\ \bibnamefont {{Bahcall}}}, \ and\
  \bibinfo {author} {\bibnamefont {et~al.}},\ }\Doi {10.1086/301513} {\bibfield
   {journal} {\bibinfo  {journal} {\aj},\ }\textbf {\bibinfo {volume} {120}},\
  \bibinfo {pages} {1579} (\bibinfo {year} {2000})},\ \Eprint
  {http://arxiv.org/abs/astro-ph/0006396} {astro-ph/0006396} \BibitemShut
  {NoStop}%
\bibitem [{\citenamefont {{Abazajian}}\ \emph {et~al.}(2009)\citenamefont
  {{Abazajian}}, \citenamefont {{Adelman-McCarthy}}, \citenamefont
  {{Ag{\"u}eros}}, \citenamefont {{Allam}}, \citenamefont {{Allende Prieto}},
  \citenamefont {{An}},\ and\ \citenamefont {et~al.}}]{abaz}%
  \BibitemOpen
  \bibfield  {author} {\bibinfo {author} {\bibfnamefont {K.~N.}\ \bibnamefont
  {{Abazajian}}}, \bibinfo {author} {\bibfnamefont {J.~K.}\ \bibnamefont
  {{Adelman-McCarthy}}}, \bibinfo {author} {\bibfnamefont {M.~A.}\ \bibnamefont
  {{Ag{\"u}eros}}}, \bibinfo {author} {\bibfnamefont {S.~S.}\ \bibnamefont
  {{Allam}}}, \bibinfo {author} {\bibfnamefont {C.}~\bibnamefont {{Allende
  Prieto}}}, \bibinfo {author} {\bibfnamefont {D.}~\bibnamefont {{An}}}, \ and\
  \bibinfo {author} {\bibnamefont {et~al.}},\ }\Doi
  {10.1088/0067-0049/182/2/543} {\bibfield  {journal} {\bibinfo  {journal}
  {\apjs},\ }\textbf {\bibinfo {volume} {182}},\ \bibinfo {pages} {543}
  (\bibinfo {year} {2009})},\ \Eprint {http://arxiv.org/abs/0812.0649}
  {arXiv:0812.0649} \BibitemShut {NoStop}%
\bibitem [{\citenamefont {{Sachs}}\ and\ \citenamefont {{Wolfe}}(1967)}]{SW}%
  \BibitemOpen
  \bibfield  {author} {\bibinfo {author} {\bibfnamefont {R.~K.}\ \bibnamefont
  {{Sachs}}}\ and\ \bibinfo {author} {\bibfnamefont {A.~M.}\ \bibnamefont
  {{Wolfe}}},\ }\Doi {10.1086/148982} {\bibfield  {journal} {\bibinfo
  {journal} {\apj},\ }\textbf {\bibinfo {volume} {147}},\ \bibinfo {pages} {73}
  (\bibinfo {year} {1967})}\BibitemShut {NoStop}%
\bibitem [{\citenamefont {{Peebles}}(1980)}]{Peeb80}%
  \BibitemOpen
  \bibfield  {author} {\bibinfo {author} {\bibfnamefont {P.~J.~E.}\
  \bibnamefont {{Peebles}}},\ }\href@noop {} {\emph {\bibinfo {title} {{The
  large-scale structure of the universe}}}}\ (\bibinfo  {publisher} {{Princeton
  University Press}},\ \bibinfo {year} {1980})\BibitemShut {NoStop}%
\bibitem [{\citenamefont {{Nusser}}\ \emph {et~al.}(2012)\citenamefont
  {{Nusser}}, \citenamefont {{Branchini}},\ and\ \citenamefont
  {{Davis}}}]{Nusser2012}%
  \BibitemOpen
  \bibfield  {author} {\bibinfo {author} {\bibfnamefont {A.}~\bibnamefont
  {{Nusser}}}, \bibinfo {author} {\bibfnamefont {E.}~\bibnamefont
  {{Branchini}}}, \ and\ \bibinfo {author} {\bibfnamefont {M.}~\bibnamefont
  {{Davis}}},\ }\Doi {10.1088/0004-637X/744/2/193} {\bibfield  {journal}
  {\bibinfo  {journal} {\apj},\ }\textbf {\bibinfo {volume} {744}},\ \bibinfo
  {eid} {193} (\bibinfo {year} {2012})},\ \Eprint
  {http://arxiv.org/abs/1106.6145} {arXiv:1106.6145 [astro-ph.CO]} \BibitemShut
  {NoStop}%
\bibitem [{\citenamefont {{Tegmark}}\ \emph {et~al.}(2004)\citenamefont
  {{Tegmark}}, \citenamefont {{Blanton}}, \citenamefont {{Strauss}},
  \citenamefont {{Hoyle}}, \citenamefont {{Schlegel}}, \citenamefont
  {{Scoccimarro}},\ and\ \citenamefont {et~al.}}]{Tegmark2004}%
  \BibitemOpen
  \bibfield  {author} {\bibinfo {author} {\bibfnamefont {M.}~\bibnamefont
  {{Tegmark}}}, \bibinfo {author} {\bibfnamefont {M.~R.}\ \bibnamefont
  {{Blanton}}}, \bibinfo {author} {\bibfnamefont {M.~A.}\ \bibnamefont
  {{Strauss}}}, \bibinfo {author} {\bibfnamefont {F.}~\bibnamefont {{Hoyle}}},
  \bibinfo {author} {\bibfnamefont {D.}~\bibnamefont {{Schlegel}}}, \bibinfo
  {author} {\bibfnamefont {R.}~\bibnamefont {{Scoccimarro}}}, \ and\ \bibinfo
  {author} {\bibnamefont {et~al.}},\ }\Doi {10.1086/382125} {\bibfield
  {journal} {\bibinfo  {journal} {\apj},\ }\textbf {\bibinfo {volume} {606}},\
  \bibinfo {pages} {702} (\bibinfo {year} {2004})},\ \Eprint
  {http://arxiv.org/abs/astro-ph/0310725} {astro-ph/0310725} \BibitemShut
  {NoStop}%
\bibitem [{\citenamefont {{Blanton}}\ \emph {et~al.}(2005)\citenamefont
  {{Blanton}}, \citenamefont {{Schlegel}}, \citenamefont {{Strauss}},
  \citenamefont {{Brinkmann}}, \citenamefont {{Finkbeiner}}, \citenamefont
  {{Fukugita}},\ and\ \citenamefont {et~al.}}]{Blanton2005}%
  \BibitemOpen
  \bibfield  {author} {\bibinfo {author} {\bibfnamefont {M.~R.}\ \bibnamefont
  {{Blanton}}}, \bibinfo {author} {\bibfnamefont {D.~J.}\ \bibnamefont
  {{Schlegel}}}, \bibinfo {author} {\bibfnamefont {M.~A.}\ \bibnamefont
  {{Strauss}}}, \bibinfo {author} {\bibfnamefont {J.}~\bibnamefont
  {{Brinkmann}}}, \bibinfo {author} {\bibfnamefont {D.}~\bibnamefont
  {{Finkbeiner}}}, \bibinfo {author} {\bibfnamefont {M.}~\bibnamefont
  {{Fukugita}}}, \ and\ \bibinfo {author} {\bibnamefont {et~al.}},\ }\Doi
  {10.1086/429803} {\bibfield  {journal} {\bibinfo  {journal} {\aj},\ }\textbf
  {\bibinfo {volume} {129}},\ \bibinfo {pages} {2562} (\bibinfo {year}
  {2005})},\ \Eprint {http://arxiv.org/abs/astro-ph/0410166} {astro-ph/0410166}
  \BibitemShut {NoStop}%
\bibitem [{\citenamefont {{Feix}}\ \emph {et~al.}(2014)\citenamefont {{Feix}},
  \citenamefont {{Nusser}},\ and\ \citenamefont {{Branchini}}}]{Feix2014}%
  \BibitemOpen
  \bibfield  {author} {\bibinfo {author} {\bibfnamefont {M.}~\bibnamefont
  {{Feix}}}, \bibinfo {author} {\bibfnamefont {A.}~\bibnamefont {{Nusser}}}, \
  and\ \bibinfo {author} {\bibfnamefont {E.}~\bibnamefont {{Branchini}}},\
  }\Doi {10.1088/1475-7516/2014/09/019} {\bibfield  {journal} {\bibinfo
  {journal} {\jcap},\ }\textbf {\bibinfo {volume} {9}},\ \bibinfo {eid} {019}
  (\bibinfo {year} {2014})},\ \Eprint {http://arxiv.org/abs/1405.6710}
  {arXiv:1405.6710} \BibitemShut {NoStop}%
\bibitem [{\citenamefont {{Blanton}}\ and\ \citenamefont
  {{Roweis}}(2007)}]{Blanton2007}%
  \BibitemOpen
  \bibfield  {author} {\bibinfo {author} {\bibfnamefont {M.~R.}\ \bibnamefont
  {{Blanton}}}\ and\ \bibinfo {author} {\bibfnamefont {S.}~\bibnamefont
  {{Roweis}}},\ }\Doi {10.1086/510127} {\bibfield  {journal} {\bibinfo
  {journal} {\aj},\ }\textbf {\bibinfo {volume} {133}},\ \bibinfo {pages} {734}
  (\bibinfo {year} {2007})},\ \Eprint {http://arxiv.org/abs/astro-ph/0606170}
  {astro-ph/0606170} \BibitemShut {NoStop}%
\bibitem [{\citenamefont {{Springel}}\ \emph {et~al.}(2005)\citenamefont
  {{Springel}}, \citenamefont {{White}}, \citenamefont {{Jenkins}},
  \citenamefont {{Frenk}}, \citenamefont {{Yoshida}}, \citenamefont {{Gao}},\
  and\ \citenamefont {et~al.}}]{Springel2005}%
  \BibitemOpen
  \bibfield  {author} {\bibinfo {author} {\bibfnamefont {V.}~\bibnamefont
  {{Springel}}}, \bibinfo {author} {\bibfnamefont {S.~D.~M.}\ \bibnamefont
  {{White}}}, \bibinfo {author} {\bibfnamefont {A.}~\bibnamefont {{Jenkins}}},
  \bibinfo {author} {\bibfnamefont {C.~S.}\ \bibnamefont {{Frenk}}}, \bibinfo
  {author} {\bibfnamefont {N.}~\bibnamefont {{Yoshida}}}, \bibinfo {author}
  {\bibfnamefont {L.}~\bibnamefont {{Gao}}}, \ and\ \bibinfo {author}
  {\bibnamefont {et~al.}},\ }\Doi {10.1038/nature03597} {\bibfield  {journal}
  {\bibinfo  {journal} {\nat},\ }\textbf {\bibinfo {volume} {435}},\ \bibinfo
  {pages} {629} (\bibinfo {year} {2005})},\ \Eprint
  {http://arxiv.org/abs/arXiv:astro-ph/0504097} {arXiv:astro-ph/0504097}
  \BibitemShut {NoStop}%
\bibitem [{\citenamefont {{Henriques}}\ \emph {et~al.}(2012)\citenamefont
  {{Henriques}}, \citenamefont {{White}}, \citenamefont {{Lemson}},
  \citenamefont {{Thomas}}, \citenamefont {{Guo}}, \citenamefont {{Marleau}},\
  and\ \citenamefont {{Overzier}}}]{Henriques2012}%
  \BibitemOpen
  \bibfield  {author} {\bibinfo {author} {\bibfnamefont {B.~M.~B.}\
  \bibnamefont {{Henriques}}}, \bibinfo {author} {\bibfnamefont {S.~D.~M.}\
  \bibnamefont {{White}}}, \bibinfo {author} {\bibfnamefont {G.}~\bibnamefont
  {{Lemson}}}, \bibinfo {author} {\bibfnamefont {P.~A.}\ \bibnamefont
  {{Thomas}}}, \bibinfo {author} {\bibfnamefont {Q.}~\bibnamefont {{Guo}}},
  \bibinfo {author} {\bibfnamefont {G.-D.}\ \bibnamefont {{Marleau}}}, \ and\
  \bibinfo {author} {\bibfnamefont {R.~A.}\ \bibnamefont {{Overzier}}},\ }\Doi
  {10.1111/j.1365-2966.2012.20521.x} {\bibfield  {journal} {\bibinfo  {journal}
  {\mnras},\ }\textbf {\bibinfo {volume} {421}},\ \bibinfo {pages} {2904}
  (\bibinfo {year} {2012})},\ \Eprint {http://arxiv.org/abs/1109.3457}
  {arXiv:1109.3457 [astro-ph.CO]} \BibitemShut {NoStop}%
\bibitem [{\citenamefont {{Nusser}}\ and\ \citenamefont
  {{Davis}}(1994)}]{Nusser1994}%
  \BibitemOpen
  \bibfield  {author} {\bibinfo {author} {\bibfnamefont {A.}~\bibnamefont
  {{Nusser}}}\ and\ \bibinfo {author} {\bibfnamefont {M.}~\bibnamefont
  {{Davis}}},\ }\Doi {10.1086/187172} {\bibfield  {journal} {\bibinfo
  {journal} {\apjl},\ }\textbf {\bibinfo {volume} {421}},\ \bibinfo {pages}
  {L1} (\bibinfo {year} {1994})},\ \Eprint
  {http://arxiv.org/abs/astro-ph/9309009} {astro-ph/9309009} \BibitemShut
  {NoStop}%
\bibitem [{\citenamefont {{Nusser}}\ \emph {et~al.}(2011)\citenamefont
  {{Nusser}}, \citenamefont {{Branchini}},\ and\ \citenamefont
  {{Davis}}}]{Nusser2011}%
  \BibitemOpen
  \bibfield  {author} {\bibinfo {author} {\bibfnamefont {A.}~\bibnamefont
  {{Nusser}}}, \bibinfo {author} {\bibfnamefont {E.}~\bibnamefont
  {{Branchini}}}, \ and\ \bibinfo {author} {\bibfnamefont {M.}~\bibnamefont
  {{Davis}}},\ }\Doi {10.1088/0004-637X/735/2/77} {\bibfield  {journal}
  {\bibinfo  {journal} {\apj},\ }\textbf {\bibinfo {volume} {735}},\ \bibinfo
  {eid} {77} (\bibinfo {year} {2011})},\ \Eprint
  {http://arxiv.org/abs/1102.4189} {arXiv:1102.4189 [astro-ph.CO]} \BibitemShut
  {NoStop}%
\bibitem [{\citenamefont {{Calabrese}}\ \emph {et~al.}(2013)\citenamefont
  {{Calabrese}}, \citenamefont {{Hlozek}}, \citenamefont {{Battaglia}},
  \citenamefont {{Battistelli}}, \citenamefont {{Bond}}, \citenamefont
  {{Chluba}},\ and\ \citenamefont {et~al.}}]{Calabrese2013}%
  \BibitemOpen
  \bibfield  {author} {\bibinfo {author} {\bibfnamefont {E.}~\bibnamefont
  {{Calabrese}}}, \bibinfo {author} {\bibfnamefont {R.~A.}\ \bibnamefont
  {{Hlozek}}}, \bibinfo {author} {\bibfnamefont {N.}~\bibnamefont
  {{Battaglia}}}, \bibinfo {author} {\bibfnamefont {E.~S.}\ \bibnamefont
  {{Battistelli}}}, \bibinfo {author} {\bibfnamefont {J.~R.}\ \bibnamefont
  {{Bond}}}, \bibinfo {author} {\bibfnamefont {J.}~\bibnamefont {{Chluba}}}, \
  and\ \bibinfo {author} {\bibnamefont {et~al.}},\ }\Doi
  {10.1103/PhysRevD.87.103012} {\bibfield  {journal} {\bibinfo  {journal}
  {\prd},\ }\textbf {\bibinfo {volume} {87}},\ \bibinfo {eid} {103012}
  (\bibinfo {year} {2013})},\ \Eprint {http://arxiv.org/abs/1302.1841}
  {arXiv:1302.1841 [astro-ph.CO]} \BibitemShut {NoStop}%
\bibitem [{\citenamefont {{Branchini}}\ \emph {et~al.}(2012)\citenamefont
  {{Branchini}}, \citenamefont {{Davis}},\ and\ \citenamefont
  {{Nusser}}}]{Branchini2012}%
  \BibitemOpen
  \bibfield  {author} {\bibinfo {author} {\bibfnamefont {E.}~\bibnamefont
  {{Branchini}}}, \bibinfo {author} {\bibfnamefont {M.}~\bibnamefont
  {{Davis}}}, \ and\ \bibinfo {author} {\bibfnamefont {A.}~\bibnamefont
  {{Nusser}}},\ }\Doi {10.1111/j.1365-2966.2012.21210.x} {\bibfield  {journal}
  {\bibinfo  {journal} {\mnras},\ }\textbf {\bibinfo {volume} {424}},\ \bibinfo
  {pages} {472} (\bibinfo {year} {2012})},\ \Eprint
  {http://arxiv.org/abs/1202.5206} {arXiv:1202.5206 [astro-ph.CO]} \BibitemShut
  {NoStop}%
\bibitem [{\citenamefont {{Song}}\ and\ \citenamefont
  {{Percival}}(2009)}]{Song2009}%
  \BibitemOpen
  \bibfield  {author} {\bibinfo {author} {\bibfnamefont {Y.-S.}\ \bibnamefont
  {{Song}}}\ and\ \bibinfo {author} {\bibfnamefont {W.~J.}\ \bibnamefont
  {{Percival}}},\ }\Doi {10.1088/1475-7516/2009/10/004} {\bibfield  {journal}
  {\bibinfo  {journal} {\jcap},\ }\textbf {\bibinfo {volume} {10}},\ \bibinfo
  {eid} {004} (\bibinfo {year} {2009})},\ \Eprint
  {http://arxiv.org/abs/0807.0810} {arXiv:0807.0810} \BibitemShut {NoStop}%
\bibitem [{\citenamefont {{Planck Collaboration}}\ \emph
  {et~al.}(2015)\citenamefont {{Planck Collaboration}}, \citenamefont {{Ade}},
  \citenamefont {{Aghanim}}, \citenamefont {{Arnaud}}, \citenamefont
  {{Ashdown}}, \citenamefont {{Aumont}}, \citenamefont {{Baccigalupi}},\ and\
  \citenamefont {et~al.}}]{Planck2015}%
  \BibitemOpen
  \bibfield  {author} {\bibinfo {author} {\bibnamefont {{Planck
  Collaboration}}}, \bibinfo {author} {\bibfnamefont {P.~A.~R.}\ \bibnamefont
  {{Ade}}}, \bibinfo {author} {\bibfnamefont {N.}~\bibnamefont {{Aghanim}}},
  \bibinfo {author} {\bibfnamefont {M.}~\bibnamefont {{Arnaud}}}, \bibinfo
  {author} {\bibfnamefont {M.}~\bibnamefont {{Ashdown}}}, \bibinfo {author}
  {\bibfnamefont {J.}~\bibnamefont {{Aumont}}}, \bibinfo {author}
  {\bibfnamefont {C.}~\bibnamefont {{Baccigalupi}}}, \ and\ \bibinfo {author}
  {\bibnamefont {et~al.}},\ }\href@noop {} {\bibfield  {journal} {\bibinfo
  {journal} {ArXiv e-prints}} (\bibinfo {year} {2015})},\ \Eprint
  {http://arxiv.org/abs/1502.01589} {arXiv:1502.01589} \BibitemShut {NoStop}%
\bibitem [{\citenamefont {{Beutler}}\ \emph {et~al.}(2012)\citenamefont
  {{Beutler}}, \citenamefont {{Blake}}, \citenamefont {{Colless}},
  \citenamefont {{Jones}}, \citenamefont {{Staveley-Smith}}, \citenamefont
  {{Poole}},\ and\ \citenamefont {et~al.}}]{Beutler2012}%
  \BibitemOpen
  \bibfield  {author} {\bibinfo {author} {\bibfnamefont {F.}~\bibnamefont
  {{Beutler}}}, \bibinfo {author} {\bibfnamefont {C.}~\bibnamefont {{Blake}}},
  \bibinfo {author} {\bibfnamefont {M.}~\bibnamefont {{Colless}}}, \bibinfo
  {author} {\bibfnamefont {D.~H.}\ \bibnamefont {{Jones}}}, \bibinfo {author}
  {\bibfnamefont {L.}~\bibnamefont {{Staveley-Smith}}}, \bibinfo {author}
  {\bibfnamefont {G.~B.}\ \bibnamefont {{Poole}}}, \ and\ \bibinfo {author}
  {\bibnamefont {et~al.}},\ }\Doi {10.1111/j.1365-2966.2012.21136.x} {\bibfield
   {journal} {\bibinfo  {journal} {\mnras},\ }\textbf {\bibinfo {volume}
  {423}},\ \bibinfo {pages} {3430} (\bibinfo {year} {2012})},\ \Eprint
  {http://arxiv.org/abs/1204.4725} {arXiv:1204.4725 [astro-ph.CO]} \BibitemShut
  {NoStop}%
\bibitem [{\citenamefont {{Samushia}}\ \emph {et~al.}(2012)\citenamefont
  {{Samushia}}, \citenamefont {{Percival}},\ and\ \citenamefont
  {{Raccanelli}}}]{Samushia2012}%
  \BibitemOpen
  \bibfield  {author} {\bibinfo {author} {\bibfnamefont {L.}~\bibnamefont
  {{Samushia}}}, \bibinfo {author} {\bibfnamefont {W.~J.}\ \bibnamefont
  {{Percival}}}, \ and\ \bibinfo {author} {\bibfnamefont {A.}~\bibnamefont
  {{Raccanelli}}},\ }\Doi {10.1111/j.1365-2966.2011.20169.x} {\bibfield
  {journal} {\bibinfo  {journal} {\mnras},\ }\textbf {\bibinfo {volume}
  {420}},\ \bibinfo {pages} {2102} (\bibinfo {year} {2012})},\ \Eprint
  {http://arxiv.org/abs/1102.1014} {arXiv:1102.1014 [astro-ph.CO]} \BibitemShut
  {NoStop}%
\bibitem [{\citenamefont {{Percival}}\ \emph {et~al.}(2004)\citenamefont
  {{Percival}}, \citenamefont {{Burkey}}, \citenamefont {{Heavens}},
  \citenamefont {{Taylor}}, \citenamefont {{Cole}}, \citenamefont {{Peacock}},\
  and\ \citenamefont {et~al.}}]{Percival2004}%
  \BibitemOpen
  \bibfield  {author} {\bibinfo {author} {\bibfnamefont {W.~J.}\ \bibnamefont
  {{Percival}}}, \bibinfo {author} {\bibfnamefont {D.}~\bibnamefont
  {{Burkey}}}, \bibinfo {author} {\bibfnamefont {A.}~\bibnamefont {{Heavens}}},
  \bibinfo {author} {\bibfnamefont {A.}~\bibnamefont {{Taylor}}}, \bibinfo
  {author} {\bibfnamefont {S.}~\bibnamefont {{Cole}}}, \bibinfo {author}
  {\bibfnamefont {J.~A.}\ \bibnamefont {{Peacock}}}, \ and\ \bibinfo {author}
  {\bibnamefont {et~al.}},\ }\Doi {10.1111/j.1365-2966.2004.08146.x} {\bibfield
   {journal} {\bibinfo  {journal} {\mnras},\ }\textbf {\bibinfo {volume}
  {353}},\ \bibinfo {pages} {1201} (\bibinfo {year} {2004})},\ \Eprint
  {http://arxiv.org/abs/astro-ph/0406513} {astro-ph/0406513} \BibitemShut
  {NoStop}%
\bibitem [{\citenamefont {{Howlett}}\ \emph {et~al.}(2015)\citenamefont
  {{Howlett}}, \citenamefont {{Ross}}, \citenamefont {{Samushia}},
  \citenamefont {{Percival}},\ and\ \citenamefont {{Manera}}}]{Howlett2014}%
  \BibitemOpen
  \bibfield  {author} {\bibinfo {author} {\bibfnamefont {C.}~\bibnamefont
  {{Howlett}}}, \bibinfo {author} {\bibfnamefont {A.~J.}\ \bibnamefont
  {{Ross}}}, \bibinfo {author} {\bibfnamefont {L.}~\bibnamefont {{Samushia}}},
  \bibinfo {author} {\bibfnamefont {W.~J.}\ \bibnamefont {{Percival}}}, \ and\
  \bibinfo {author} {\bibfnamefont {M.}~\bibnamefont {{Manera}}},\ }\Doi
  {10.1093/mnras/stu2693} {\bibfield  {journal} {\bibinfo  {journal} {\mnras},\
  }\textbf {\bibinfo {volume} {449}},\ \bibinfo {pages} {848} (\bibinfo {year}
  {2015})},\ \Eprint {http://arxiv.org/abs/1409.3238} {arXiv:1409.3238}
  \BibitemShut {NoStop}%
\bibitem [{\citenamefont {{Tully}}\ and\ \citenamefont
  {{Fisher}}(1977)}]{TF77}%
  \BibitemOpen
  \bibfield  {author} {\bibinfo {author} {\bibfnamefont {R.~B.}\ \bibnamefont
  {{Tully}}}\ and\ \bibinfo {author} {\bibfnamefont {J.~R.}\ \bibnamefont
  {{Fisher}}},\ }\href@noop {} {\bibfield  {journal} {\bibinfo  {journal}
  {\aap},\ }\textbf {\bibinfo {volume} {54}},\ \bibinfo {pages} {661} (\bibinfo
  {year} {1977})}\BibitemShut {NoStop}%
\bibitem [{\citenamefont {{Macaulay}}\ \emph {et~al.}(2013)\citenamefont
  {{Macaulay}}, \citenamefont {{Wehus}},\ and\ \citenamefont
  {{Eriksen}}}]{Macaulay2013}%
  \BibitemOpen
  \bibfield  {author} {\bibinfo {author} {\bibfnamefont {E.}~\bibnamefont
  {{Macaulay}}}, \bibinfo {author} {\bibfnamefont {I.~K.}\ \bibnamefont
  {{Wehus}}}, \ and\ \bibinfo {author} {\bibfnamefont {H.~K.}\ \bibnamefont
  {{Eriksen}}},\ }\Doi {10.1103/PhysRevLett.111.161301} {\bibfield  {journal}
  {\bibinfo  {journal} {Physical Review Letters},\ }\textbf {\bibinfo {volume}
  {111}},\ \bibinfo {eid} {161301} (\bibinfo {year} {2013})},\ \Eprint
  {http://arxiv.org/abs/1303.6583} {arXiv:1303.6583 [astro-ph.CO]} \BibitemShut
  {NoStop}%
\bibitem [{\citenamefont {{Levi}}\ \emph {et~al.}(2013)\citenamefont {{Levi}},
  \citenamefont {{Bebek}}, \citenamefont {{Beers}}, \citenamefont {{Blum}},
  \citenamefont {{Cahn}}, \citenamefont {{Eisenstein}},\ and\ \citenamefont
  {et~al.}}]{Levi2013}%
  \BibitemOpen
  \bibfield  {author} {\bibinfo {author} {\bibfnamefont {M.}~\bibnamefont
  {{Levi}}}, \bibinfo {author} {\bibfnamefont {C.}~\bibnamefont {{Bebek}}},
  \bibinfo {author} {\bibfnamefont {T.}~\bibnamefont {{Beers}}}, \bibinfo
  {author} {\bibfnamefont {R.}~\bibnamefont {{Blum}}}, \bibinfo {author}
  {\bibfnamefont {R.}~\bibnamefont {{Cahn}}}, \bibinfo {author} {\bibfnamefont
  {D.}~\bibnamefont {{Eisenstein}}}, \ and\ \bibinfo {author} {\bibnamefont
  {et~al.}},\ }\href@noop {} {\bibfield  {journal} {\bibinfo  {journal} {ArXiv
  e-prints}} (\bibinfo {year} {2013})},\ \Eprint
  {http://arxiv.org/abs/1308.0847} {arXiv:1308.0847 [astro-ph.CO]} \BibitemShut
  {NoStop}%
\bibitem [{\citenamefont {{Laureijs}}\ \emph {et~al.}(2011)\citenamefont
  {{Laureijs}}, \citenamefont {{Amiaux}}, \citenamefont {{Arduini}},
  \citenamefont {{Augu{\`e}res}}, \citenamefont {{Brinchmann}}, \citenamefont
  {{Cole}},\ and\ \citenamefont {et~al.}}]{euclid2011}%
  \BibitemOpen
  \bibfield  {author} {\bibinfo {author} {\bibfnamefont {R.}~\bibnamefont
  {{Laureijs}}}, \bibinfo {author} {\bibfnamefont {J.}~\bibnamefont
  {{Amiaux}}}, \bibinfo {author} {\bibfnamefont {S.}~\bibnamefont {{Arduini}}},
  \bibinfo {author} {\bibfnamefont {J.~.}\ \bibnamefont {{Augu{\`e}res}}},
  \bibinfo {author} {\bibfnamefont {J.}~\bibnamefont {{Brinchmann}}}, \bibinfo
  {author} {\bibfnamefont {R.}~\bibnamefont {{Cole}}}, \ and\ \bibinfo {author}
  {\bibnamefont {et~al.}},\ }\href@noop {} {\bibfield  {journal} {\bibinfo
  {journal} {ArXiv e-prints}} (\bibinfo {year} {2011})},\ \Eprint
  {http://arxiv.org/abs/1110.3193} {arXiv:1110.3193 [astro-ph.CO]} \BibitemShut
  {NoStop}%
\end{thebibliography}%
\end{document}